# A Robust and Tunable Add-Drop Filter using Whispering Gallery Mode Microtoroid Resonator

Faraz Monifi, Jacob Friedlein, Sahin Kaya Ozdemir, and Lan Yang,

*Abstract*—We fabricated and theoretically investigated an add-drop filter using an on-chip whispering gallery mode (WGM) microtoroid resonator with ultra-high quality factor ($Q$) side coupled to two taper fibers, forming the bus and drop waveguides. The new device design incorporates silica side walls close to the microresonators which not only enable placing the coupling fibers on the same plane with respect to the microtoroid resonator but also provides mechanical stability, leading to an add-drop filter with high drop efficiency and improved robustness to environmental perturbations. We show that this new device can be thermally tuned to drop desired wavelengths from the bus without significantly affecting the drop efficiency, which is around 57%.

*Index Terms*—Whispering gallery mode (WGM), microresonator, add-drop filter, microtoroid, thermal effects

## I. INTRODUCTION

PTICAL add-drop filters (ADFs) that add or remove narrow band wavelengths of light from a broader optical signal being carried along a bus waveguide are fundamental building blocks of optical transmission and communication systems. They are key elements in multiplexers, modulators and optical switches. Thus, great efforts have been put in designing ADF architectures with improved efficiency, spectral selectivity (i.e., quality factor), and spectral tunability. The past two decades have seen significant progress in ADF designs ranging from all-fiber architectures [1]–[5] to photonic crystal (PhC) structures [6]–[8] and waveguide coupled whispering gallery mode (WGM) microresonators (e.g., microsphere, microring, microdisk, microtoroid, etc) [9]–[15].

All-fiber systems relying on Bragg gratings accomplish high drop efficiency but have quality factors of only about $\sim 10^3$, which yields a wavelength selectivity of about 0.7nm (or equivalently a frequency selectivity of $\sim 87.7$GHz) [2]–[5]. Another all-fiber scheme using a fiber taper coupled microfiber knot, which works as a resonator, has been reported to have a quality factor of $Q \sim 1.3 \times 10^4$ [1]. In a two-dimensional PhC with a complete photonic band-gap, nearly 100% drop efficiency and a $Q$ close to 700 have been numerically demonstrated, however, with a very small overall transmission ($\sim 15\%$ of the incident power) [6]. A proposal using PhC ring resonator has promised more than 96% drop efficiency with $160 < Q < 10^3$. These PhC based ADFs are fabricated with very high index materials which are considered not compatible with the low-index materials such as silica used in optical telecommunication systems [7], [8].

Add-drop filters using WGM microresonators have shown great promise for practical applications as such resonators have very high quality factors (up to $10^9$) [16] and can be coupled to optical waveguides commonly used in optical communications with efficiencies up to 99.7% [17]. Among all WGM microresonators, microrings have been studied most thoroughly due to their ease of fabrication and on chip structure [9]–[12]. Although microring ADFs are very promising, they face several drawbacks. For instance, fabrication imperfections limit their $Q$ to $10^6$, and their performance is highly sensitive to the nanoscale gap between the ring resonator and the waveguides, which imposes a manufacturing challenge. Moreover, they suffer from coupling losses, because they are not fiber based structures. In these structures, light carried in the standard optical communication network via fibers has to be first coupled into a waveguide (not fiber or tapered fiber) built on the same chip as the resonator to couple light in and out of the ADF, and the light in the drop or the throughput waveguide is then coupled back to the fiber optical network. This increases the losses.

In order to circumvent the fiber-waveguide coupling related problems in microring ADFs while making use of the high-$Q$ of WGM resonators, ADFs using fiber taper coupled microspheres [13], [14] and microtoroid resonators were proposed and fabricated. Drop efficiencies of about 50% and $Q > 10^6$ have been reported for fiber taper coupled microsphere-based ADFs [13]. Use of microspheres as ADFs is limited by the fact that identical microspheres cannot be mass produced and they are not on-chip devices; rather, they are suspended at the tips of fibers since a microsphere is usually fabricated by heating the tip of a tapered fiber under flame or $CO_2$ laser. In addition, the coupling fiber tapers are also suspended. This leads to mechanical stability problems. Moreover, the relatively large mode volumes of microspheres leads to many closely spaced resonant modes [18], resulting in a small free spectral range, which is undesirable for ADF applications because a small free spectral range implies a smaller bandwidth.

On-chip design, ultra-high-Q, microscale mode volume, large free spectral range [19], high coupling efficiency and ease of integration into existing optical fiber networks make fiber taper coupled silica microtoroid resonators good candidates for ADFs. Recently, Yao et al demonstrated an on-chip add-drop filter using a silicon microtoroid with MEMS actuated bus and drop waveguides moving the microtoroid based ADFs a step closer to practicality [20], [21]. However, the demonstrated intrinsic quality factors of $Q \sim 10^5$ is much lower than that obtained with silica microtoroids with fiber couplers and have the same coupling losses (explained above) as the silicon microring based ADFs. Previously, ADFs with quality factors of $3 \times 10^6$ and close to 94% drop efficiency have been reported [15]. However, the coupling fiber tapers are still suspended making it difficult to align the fibers on the same plane as the microtoroid, and causing mechanical instabilities and fluctuations in the gap between the tapers and the resonator.

In a study of optomechanics in WGM resonators, Lin *et al.* have shown that a pair of nanoforks fabricated near a microdisk resonator forms a good support to provide mechanical stability to a fiber taper coupled to the microdisk [22]. The reported total insertion loss is ∼ 8%. The nanofork concept has not been used in microtoroid-taper coupling process and certainly not in microdisk or microtoroid based ADFs, which require two pairs of such nanoforks fabricated on two opposite sides of the resonator. Here, we present a new low-loss (∼ 1%) and easy-to-fabricate device, which addresses mechanical stability problem in fiber taper coupled resonators, in particular in ADF schemes, and provides remedies to enhance the performance of microtoroid-based ADFs.

The new device we introduce in this paper consists of a reflowed silica sidewall fabricated on one side of an array of on-chip microtoroid resonators to form a support for fiber taper waveguides, providing mechanical stability and improving the performance of microtoroid-based ADFs. The sidewall has a very low insertion loss, and enables the alignment of the fiber-tapers on the same plane as the microtoroid much easier and more robust. Applying this new concept, we achieved a drop efficiency of 57% with a loaded Q-factor of $4.5 \times 10^6$, which is higher than that in previously reported ADFs. We demonstrated the thermal tunability of our device, and showed the drop wavelength could be tuned within a band of $30 GHz$ by changing the temperature in the range of $20-34^{o}C$ without significant change in the drop efficiency.

The paper is organized as follows. Section II includes theoretical background necessary to understand the physics of WGM resonators and the working principle of ADFs, as well as criteria for the evaluation of ADF performance. In Sec. III, we first introduce the new design and the fabrication process of our device and then report the experimental results in detail. Finally, Sec. IV includes conclusions and discussion of the results.

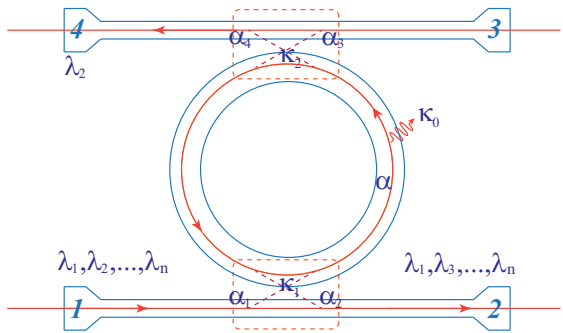

Fig. 1. Illustration of an ADF. A channel (signal) with wavelength $\lambda_2$ of input mode $a1$ at port 1 of a bus waveguide is dropped at port 4 (drop mode: $a_4$) by the ADF fabricated using a whispering gallery mode (WGM) resonator having a resonance at $\lambda_2$. The field mode at the throughput port 2 is denoted by mode $a_2$, and the intra-cavity field mode is $a$. Port 3 with mode $a_3$ is the add port of the add-drop waveguide. Waveguide-resonator coupling losses are denoted as $k_1$ and $k_2$, and the intrinsic cavity losses are denoted by $k_0$. In this illustration port 3 (add port) is left at vacuum (add mode: $a_3= 0$).

## II. THEORETICAL BACKGROUND

### A. Drop efficiency and transmission

Figure 1 shows a schematic representation of a WGM microresonator based ADF. The resonator is evanescently side coupled to a pair of tapered fibers, one of which is used as the bus waveguide which includes the input port for the field $a1$ contains a series of wavelengths, $\lambda_1, \lambda_2, ...., \lambda_n$ while the other fiber is used as the drop waveguide and includes the add and drop ports for the fields $a3$ and $a4$, respectively. The signal at the input port having wavelength $\lambda_i$ which is at resonance with the microresonator, will first be coupled into the resonator creating the normalized cavity field mode $a$ (i.e., $|a|_2$ is the energy stored in the cavity) and then will be transferred to the drop port $a_4$, whereas signals with non-resonant wavelengths are transmitted to the throughput port $a_2$ without loss. Similarly, a signal at the add port $a_3$ with a wavelength at resonance with the resonator will be transmitted to the port $a_2$ via the microresonator, joining the non-resonant wavelengths of the input which pass to $a_2$ with minimal loss.

Denoting the resonator total round-trip intrinsic loss, the bus-resonator coupling loss and the drop-resonator coupling loss as $k_0, k_1$ and $k_2$, respectively, we can write the time evolution of the intra-cavity field $a$ of the microresonator as

$$\frac{da}{dt} = -\left(j\omega_c + \frac{\kappa_0 + \kappa_1 + \kappa_2}{2}\right)a - \sqrt{\kappa_1}a_1 - \sqrt{\kappa_2}a_3 \tag{1}$$

where $\omega_c$ is the resonance frequency of the resonator. Input-output relations for the through and the drop ports are respectively described by

$$a_2 = a_1 + \sqrt{\kappa_1}a \tag{2}$$

and

$$a_4 = a_3 + \sqrt{\kappa_2}a \tag{3}$$

respectively. For simplicity, let us assume that $a_3 = 0$, that is there is no input at the add port. Then assuming the intra-cavity field is a single wavelength field with frequency $\omega$ and defining the cavity-field detuning as, $\Delta = \omega - \omega_c$ we find from Eqs.1 and 2 that at steady-state the input $a_1$ and the throughput $a_2$ satisfy

$$\left(j\Delta - \frac{\kappa_0 + \kappa_1 + \kappa_2}{2}\right)(a_2 - a_1) = \kappa_1 a_1 \tag{4}$$

from which we find the amplitude transmission coefficient $t$ as

$$t = \frac{a_2}{a_1} = \frac{j2\Delta - (\kappa_0 - \kappa_1 + \kappa_2)}{j2\Delta - (\kappa_0 + \kappa_1 + \kappa_2)}, \tag{5}$$

and subsequently the power transmission coefficient $T = |t|^2$ as

$$T = \frac{4\Delta^2 + (\kappa_0 - \kappa_1 + \kappa_2)^2}{4\Delta^2 + (\kappa_0 + \kappa_1 + \kappa_2)^2}. \tag{6}$$

Then at resonance $\Delta = 0$, transmission coefficient becomes

$$T = \frac{(\kappa_0 - \kappa_1 + \kappa_2)^2}{(\kappa_0 + \kappa_1 + \kappa_2)^2}. \tag{7}$$

Recalling that $T = 0$ at critical coupling, we find the critical coupling condition for the add-drop configuration as $\kappa_1 = \kappa_0 + \kappa_2$. Note that in standard configurations where the coupling between a taper and a resonator is considered, critical coupling is described $\kappa_1 = \kappa_0$, that is the coupling loss equals the intrinsic loss. In the ADF configuration, the difference comes from the second coupling taper which introduces an additional loss described by $\kappa_2$. When we look at the system from the bus side, as is the case when there is only one coupling taper, this extra loss induced by the drop waveguide, may be regarded as an increase in the intrinsic loss of the resonator. Thus, fundamentally the situation is similar, that is the critical coupling for the ADF configuration is achieved when the intrinsic loss equals the coupling loss of the bus. Figure 2 depicts the dependence of throughput transmission on $\kappa_1$.

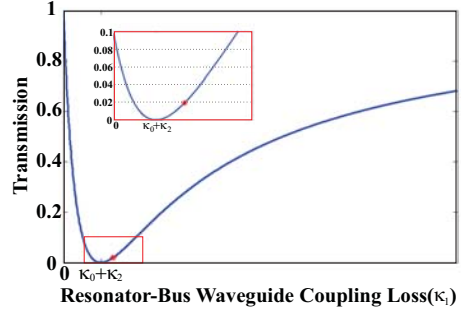

Fig. 2. Simulation results showing the dependence of the transmission (signal at the throughput port) as a function of the bus waveguide-resonator coupling loss $\kappa_1$ when the sum of the add-drop waveguide resonator loss $\kappa_2$ and the intrinsic loss $\kappa_0$ of the cavity is kept constant. When $\kappa_1 = \kappa_0 + \kappa_2$, transmission is zero corresponding to critical coupling for the bus waveguide. The point labeled as * denotes the experimental condition used in this study.

Amplitude drop efficiency of the ADF, which is defined as *d* can be calculated using Eqs. 1 and 3 as

$$d = \frac{a_4}{a_1} = \frac{2\sqrt{\kappa_1 \kappa_2}}{j2\Delta - (\kappa_0 + \kappa_1 + \kappa_2)}. \tag{8}$$

Subsequently, power drop efficiency which is defined as the ratio of the power transferred to the drop port to the input power, $D = |d|^2 = |a_4/a_1|^2$ is found as

$$D = \frac{4\kappa_1 \kappa_2}{4\Delta^2 + (\kappa_0 + \kappa_1 + \kappa_2)^2} \tag{9}$$

which becomes

$$D = \frac{4\kappa_1 \kappa_2}{(\kappa_0 + \kappa_1 + \kappa_2)^2}. \tag{10}$$

at resonance $\Delta = 0$. For an ideal ADF, the signal to be dropped should be transferred to the drop port with unit efficiency and transmission to the throughput for this signal should be zero. The latter is satisfied at critical coupling condition derived above as $\kappa_1 = \kappa_0 + \kappa_2$. Thus, we need to maximize Eq. 10 with the constraint that $\kappa_1 = \kappa_0 + \kappa_2$. Substituting this into Eq. 10 we arrive at

$$D = \frac{\kappa_2}{\kappa_0 + \kappa_2} = \frac{\kappa_2}{\kappa_1} = \frac{1}{1 + \kappa_0/\kappa_2} = 1 - \frac{\kappa_0}{\kappa_1}. \tag{11}$$

Note that $\kappa_0$ is fixed with the choice of the resonator material, size, geometry and fabrication quality, therefore, one cannot tune it once the resonator is fabricated. However, one can easily modify the coupling losses $\kappa_1$ and $\kappa_2$ by changing the coupling distance (air gap) between the tapers and the resonator. Equation 11 suggests the add-drop waveguide should be brought to deep over-coupling regime, where $\kappa_2 \gg \kappa_0$, to achieve $D \simeq 1$. Then, $\kappa_1$ should be set to satisfy $\kappa_1 = \kappa_0 + \kappa_2$ (i.e., $\kappa_1 \sim \kappa_2$) for zero transmission in the throughput. This, however, implies that bus waveguide should also be set to over-coupling region, because $\kappa_2$ cannot be increased arbitrarily without increasing $\kappa_1$ due to the critical coupling condition which implies that increasing $\kappa_2$ increases $\kappa_1$, too. On the other hand, if we initially set $\kappa_0 = \kappa_2$, then $D$ becomes $D = 1/2$. In Fig. 3, we plot the drop efficiency $D$ as a function of $\kappa_1$ and $\kappa_2$ for fixed $\kappa_0$. It can be seen that the maximum occurs when $\kappa_1 = \kappa_0 + \kappa_2$, and increasing $\kappa_2/\kappa_0$ increases the maximum value that $D$ can attain. Maximum drop efficiency obtained from the zero transmission condition is equivalent to maximization of Eq. 10 with respect to $\kappa_1$ while $\kappa_0$ and $\kappa_2$ are kept constant. One may also try to maximize Eq. 10 with respect to $\kappa_2$ while keeping $\kappa_0$ and $\kappa_1$ constant. In doing so, one derives the condition $\kappa_2 = \kappa_0 + \kappa_1$ which leads to

$$D' = \frac{\kappa_1}{\kappa_0 + \kappa_1} = \frac{\kappa_1}{\kappa_2} = \frac{1}{1 + \kappa_0/\kappa_1}. \quad (12)$$

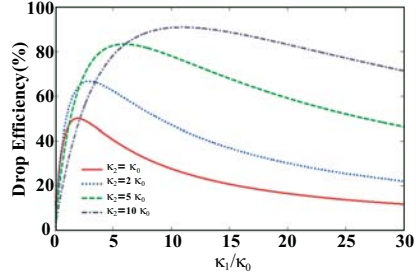

Fig. 3. Simulation results showing the dependence of drop efficiency $D$ on the resonator intrinsic loss $\kappa_0$, waveguide-resonator coupling loss $\kappa_1$ and the add-drop waveguide resonator coupling loss $\kappa_2$. For a resonator of loss $\kappa_0$ there is a ($\kappa_1$, $\kappa_2$) pair which maximizes the drop efficiency. For a given $\kappa_2$, increasing $\kappa_2$ increases the maximum $D$ and shifts $\kappa_1$ to higher values.

In this case, from Eq. 7 we find the transmission as

$$T' = \frac{\kappa_0^2}{(\kappa_0 + \kappa_1)^2} = \frac{1}{(1 + \kappa_1/\kappa_0)^2}. \quad (13)$$

Equation 12, too, implies that $D' \simeq 1$ if the bus waveguide is set to over-coupling, i.e., $\kappa_1 \gg \kappa_0$, the add-drop waveguide must be over-coupled too, to satisfy $\kappa_2 = \kappa_0 + \kappa_1$. This then results in $T' \simeq 0$ as desired. Thus, both maximization approaches lead to the same result, that is both the add-drop and the bus resonator coupling should be brought to over-coupling regime. Now let us assume that before the add-drop fiber taper is brought to the system, the bus fiber taper and the resonator is set to critical coupling ($\kappa_0 = \kappa_1$) where we expect the transmission to be zero. If we set $\kappa_0 = \kappa_1$ in Eq. 13, we find $T' = 1/4$ and $D' = 1/2$, implying that with the introduction of the second waveguide the critical coupling condition for the first waveguide-resonator coupling is modified, which in turn, shifts the drop efficiency and the transmission away from their optimal values. Thus, the loss induced by the second waveguide should be taken into account. Similarly, if we initially set the bus and the resonator in deep under-coupling $\kappa_0 \gg \kappa_1$, then we are forced to choose $\kappa_0 \sim \kappa_2$, and we lose our ability to tune that coupling, thus ending with $T' \simeq 1$ and $D' \simeq 0$.

Another issue that should be kept in mind when designing an ADF is the wavelength selectivity of the device that is the quality factor $Q$ of the device, which depends strongly on intrinsic loss and coupling losses. Quality factor $Q$ is defined as

$$\frac{1}{Q} = \frac{1}{Q_0} + \frac{1}{Q_1} + \frac{1}{Q_2} \tag{14}$$

where $Q_0 = \omega/\kappa_0$, $Q_1 = \omega/\kappa_1$ and $Q_2 = \omega/\kappa_2$ are the intrinsic quality factor, bus-resonator coupling quality factor, and add-drop resonator coupling quality factor, respectively. Increasing $\kappa_2$ and/or $\kappa_1$ leads to a decrease in $Q$, which in turn, deteriorates the wavelength selectivity of the ADF. This implies a trade-off between $D$, $T$ and $Q$.

Finally, in an ADF filter, power conservation requires that $P_{\text{drop}} + P_{\text{transmitted}} + P_{\text{loss}} = P_{\text{in}}$. From Eqs. 9 and 12, one can show that normalized loss $L$ is given by

$$L = \frac{P_{\text{loss}}}{P_{\text{in}}} = \frac{4\kappa_0 \kappa_1}{(\kappa_0 + \kappa_1 + \kappa_2)^2} = \frac{\kappa_0}{\kappa_2} D \tag{15}$$

from which we obtain

$$1 - T = D\left(1 + \frac{\kappa_0}{\kappa_2}\right) = D\left(1 + \frac{Q_2}{Q_0}\right). \tag{16}$$

*B. Group delay in add-drop filters*

Waveguide coupled high-$Q$ micro-resonators show strong normal and anomalous dispersion near the resonances due to the rapid variation in the effective phase of the field. This, in turn, leads to slow and fast light phenomena, which affect the response time of ADFs fabricated using resonators. In order to quantify this, we return to the expressions for $t$ and $d$ given in Eqs.5 and 8 and calculate the effective phase shift $\Phi$ and the group delay $\tau$ for the throughput and drop fields.

Recalling that $\Phi_t = \arg t$ and the group delay $\tau_t = d\Phi_t/d\Delta$, we find

$$\Phi_t = \arctan\frac{\Delta \kappa_1}{\Delta^2 + \xi} \tag{17}$$

where we have used $\xi = \left[(\kappa_0 + \kappa_2)^2 - \kappa_1^2\right]/4$. Subsequently, we calculate the group delay for the bus waveguide as

$$\tau_t = \frac{\kappa_1(\Delta^2 - \xi)}{\Delta^2\kappa_1^2 + (\Delta^2 + \xi)^2} \tag{18}$$

which becomes

$$\lim_{\Delta \to 0} \tau_t = \frac{-\kappa_1}{\xi} = \frac{-4\kappa_1}{(\kappa_0 + \kappa_2)^2 - \kappa_1^2} \tag{19}$$

around the resonance (i.e., zero detuning, $\Delta \to 0$). Thus, for the bus waveguide we observe slow light for $\kappa_1 > \kappa_0 + \kappa_2$ and fast light for $\kappa_1 < \kappa_0 + \kappa_2$. Note that zero transmission is achieved when $\kappa_1 = \kappa_0 + \kappa_2$ for which neither the slow nor fast light phenomenon is relevant. However, slight deviations from this critical coupling condition will lead to non-zero transmission which may be slower or faster depending on whether the ADF is operated in the over-coupling or under-coupling regime.

Next, we carry out similar steps for the drop signal and find $\Phi_d = \arg d$ and the group delay $\tau_d = d\Phi_d/d\Delta$ as

$$\Phi_d = \arctan\frac{-2\Delta}{\kappa_0 + \kappa_1 + \kappa_2} = \arctan\frac{2\Delta}{\chi} \tag{20}$$

and

$$\tau_d = \frac{2\chi}{\chi^2 + 4\Delta^2} \tag{21}$$

where $\chi = \kappa_1 + \kappa_0 + \kappa_2$. At the resonance, the group delay $\tau_d$ becomes

$$\lim_{\Delta \to 0} \tau_d = \frac{2}{\chi} = \frac{2}{\kappa_0 + \kappa_1 + \kappa_2} \tag{22}$$

which is always positive (i.e., $\tau_d > 0$) because the coupling losses and the intrinsic loss for this passive resonator based ADF are always positive. Hence, the signal at the drop experiences group delay and is always in the slow light regime. Inserting the zero-transmission condition in Eq.22, we find the group delay of the drop signal as

$$\lim_{\Delta \to 0} \tau_d = \frac{1}{\kappa_1} \tag{23}$$

implying that decreasing the group delay requires increased coupling losses on the bus waveguide. However, this will decrease the $Q$ of the ADF as we have discussed above.

We conclude this subsection by noting that in the ADF design, there is a tradeoff between the drop efficiency, quality factor and the group delay experienced by the dropped signal. Thus, one needs to optimize these parameters for the specific application of interest.

## III. EXPERIMENTS

In this section, we introduce the new design of chip and the fabrication of the ADF using microtoroid resonators with reflowed silica walls for the stabilization of taper fibers, which form the bus and add-drop waveguides, and then we present the experimental results discussing the performance and thermal tuning of the device. Fig. 4.

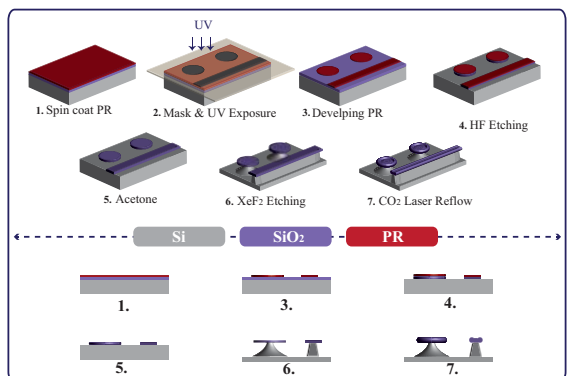

Fig. 4. Illustration showing the fabrication process with the top (upper panel) and side (lower panel) views of final product after each fabrication step. See text for details. PR: Photoresist, UV: Ultraviolet, HF: Hydrofluoric acid, $XeF_2$: Xenon difluoride, Si: Silicon, $SiO_2$: Silica and $CO_2$: Carbon Dioxide

### A. Fabrication of microtoroids with close side walls

Microtoroids are fabricated from silica-on-silicon wafers through a three step process as established by Armani *et al.* [19] (i) Standard photolithography for patterning circular silica disks on silicon wafer using a mask composed of rows of circles of various sizes. (ii) Selective isotropic etching of silicon using xenon difluoride ($XeF_2$) to form undercut structure with silica disks over silicon pillars. (iii) $CO_2$ reflow to transform the disk to a torus forming a silica microtoroid on a silicon pillar. At the $CO_2$ laser wavelength of ~ 10.6$\mu m$, the absorption cross-section of silica is much larger than that of silicon. Thus, the silicon pillar acts as a heat sink transferring the heat to the silicon substrate, while silica absorbs the energy and is softened such that surface tension rolls up the softened silica edges forming a toroidal shape.

The fabrication process of our new design differs from that of the standard process (i)-(iii) of toroid fabrication in two ways. First, the mask used for photolithography has a straight line along each row of circles, which will serve as the side wall for the stabilization of the taper fibers and for the ease of alignment of the two fiber taper waveguides and the resonator in the same plane. Thus, after steps (i) and (ii), we obtain a series of undercut silica disks over silicon pillars and undercut silica-on-silicon rectangular walls spanning from one end to the other end of the wafer along the rows of disks. Second $CO_2$ reflow of the walls. After performing step (iii), we perform step (iv): $CO_2$ reflow of the rectangular walls to transform the rectangular walls with rough edges into smoother rolled up edges via the mechanism explained in (iii). Figure 4 depicts the fabrication steps of the proposed device and side views of the structure for each fabrication step. Using the aforementioned process, we fabricated microtoroid resonators of major diameter $100-120\mu m$ and minor diameter $9-10\mu m$ with side walls of $300\mu m$ width and $70-80\mu m$ height. In one design, we had a single continuous wall spanning along the row of microtoroids. In the other, we had a shorter wall next to each toroid. Thanks to the reflow process, which prepares very smooth rolled-up edges, the fiber taper waveguides have very small contact areas with the sidewall surface. This minimizes scattering losses, leading to a measured insertion loss of ~ 1%. In Fig. 5, we show an illustration of an array of microtoroids with a sidewall to support two fiber-taper waveguides, as well as SEM images of the fabricated devices.

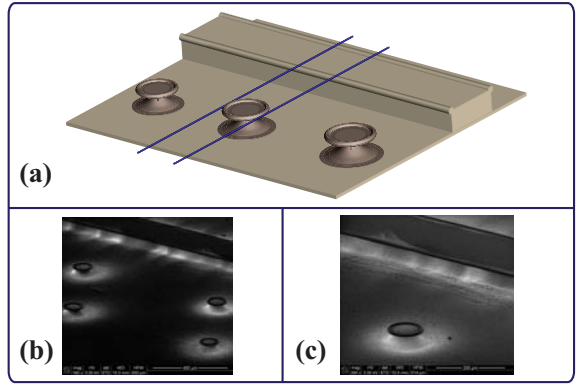

Fig. 5. (a) A 3D rendering image of an array of toroids with a side wall to support two fiber-taper waveguides which serve as bus and drop channels, respectively. The side wall made of silica is reflowed by a $CO_2$ to minimize scattering loss. (b) Scanning electron microscope (SEM) image of an array of microtoroid resonators and a side wall. (c) Magnified image of one of the resonators shown in (b) together with its side wall.

One may consider placing a second reflowed silica sidewall on the other side of the resonator to increase mechanical stability. However, we observed that such a second sidewall induces extra losses and hence decreases the drop efficiency, because in this case if we trace the signal path from input to drop port, we encounter two contact regions (i.e., one for each waveguide), each of which induces ~ 1% loss. In our experiments, we could not observe a significant improvement in mechanical stability or alignment process when two sidewalls were used; therefore, we used one sidewall configuration to keep the fabrication process simple and losses small.

### B. Fabrication of waveguides

In this work, we use fiber-tapers as bus and add-drop waveguides. Using these fiber tapers rather than other types of waveguides makes it easier to integrate the constructed ADFs to the existing optical communication networks. We fabricated the fiber tapers from standard single mode communication fibers by heat-and-pull method over hydrogen flame. For the performance of the ADF, it is crucial that the two fiber-tapers are precisely positioned along two sides of the microtoroid resonator. To obtain two parallel tapers with a separation on the order of a few tens of micrometers, we placed two fibers in parallel in a holder and then heated and pulled them simultaneously. In principle, this should result in two uniform fiber tapers with nearly the same size and the same mode profile. After the pulling, the fiber tapers are usually loose, we tighten them to reduce the bending caused by loose fibers (This will help us to improve the vertical alignment of the fiber tapers with microtoroid). The tapers had waist diameters of a few microns and exhibited about 90% transmission, which is lower than what is typically achieved when a single fiber taper is pulled. This discrepancy may be attributed to the difficulty in optimizing the heating conditions for both fibers at the same time for the pulling process. Moreover, although the tapers were pulled in parallel, there was a clear height difference in the final product. Pulling two fibers simultaneously allows us to place two fiber-tapers closer to each other, but it does not solve the problem of fine-tuning the horizontal and vertical positions of tapers with respect to each other and the toroid. The walls fabricated along the resonators provide a remedy for this problem (see Fig. 6).

### C. Alignment and assembly of add-drop filter

We placed the chip containing the microtoroids on a 3D nano-positioning system to precisely tune its position with respect to the parallel tapers. We first raised the chip to side couple the tapers to the resonator without the help of the side walls. Figure 6 shows the optical microscope images of the alignment process. As the chip was raised, the microtoroid of interest entered the focal plane of the microscope. We adjusted both the height and horizontal position of the resonator so that it was on the same plane and side coupled with one of the tapers. During this process the other taper was out of sight (see Fig. 6 top panel). The side view with a microscope reveals the height difference between the tapers. Then we slowly changed the focal point of the microscope which pushed the first taper fiber out of focus and the second taper started to come into sight. We achieved in bringing this second fiber into the focus, however, the first fiber taper was not at the same plane (see Fig. 6 middle panel) and the resonator could not be resolved. All fiber taper pairs fabricated in our experiments suffered from this height difference which made it difficult to align the tapers with the resonator in the same plane.

Next, we used the fabricated side walls with smooth rolled up edges to solve the aforementioned problem. We first roughly adjusted the horizontal distance between the tapers by pushing one of the tapers towards the other using a fiber tip controlled with a micro-positioning stage until the distance between the tapers is roughly the diameter of the resonator. The tapers can be pushed towards to or away from each other and the microtoroid resonator depending on which side of the waveguide the fiber tip is positioned. Then we raised the chip carefully until the lower taper rested on the wall. We continued raising the chip, which also raised the toroid and the aligned lower taper, until both tapers were at the same height and in the same plane as the toroid. Then with the help of the 3D nano-positioning system and the fiber tip, we fine-tuned the distance between the fiber tapers as well as the distance between the tapers and the microtoroid such that near-optimal coupling to the resonator was achieved. Since both fibers rested on the wall, fine-tuning the horizontal distances using the fiber tip and the nano-positioning system did not change their heights. When the satisfactory coupling is achieved, the fiber tip was removed. Since the taper fiber waveguides rested on the sidewall, their positions did not change upon the removal of the fiber tip. Note that if the fiber taper waveguides had not been on the sidewall, they would have returned close to their initial pre-alignment positions upon the removal of the fiber tip due to restoring forces.

We should note here that the fine tuning with fiber tip induces stress on the fiber taper. However, the taper stays in its position as long as there is no external perturbation such as thermally induced extra stress or strong mechanical fluctuations of the setup. For out of the lab applications, one may isolate the add-drop filter from the surrounding to minimize such perturbations.

Top and side views of the final alignment shown in the bottom panel of Fig. 6 show that the taper fibers have reached almost the same height and are aligned with each other and with the microtoroid, thus clearly demonstrating the improvement over the process without the walls. The accuracy of the vertical alignment depends on the evenness of the height of the sidewall at the contact regions. Slight height variations along the sidewall may be caused by uneven reflow. If the reflow is sufficiently even, variations will be negligible and vertical alignment will be near-perfect. The fact that we were able to obtain drop signals from multiple different samples indicates that we can repeatedly produce sidewalls without significant height variations, and that slight vertical misalignments do not cause a significant problem. With a much better operator care and strong $CO_2$ laser beam for the reflow, accuracy of the vertical alignment can be easily improved.

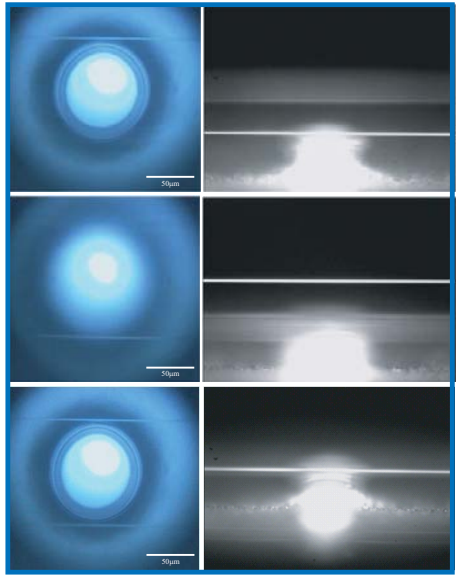

Fig. 6. Top (left column) and side (right column) views obtained with optical microscope during the alignment of fiber tapers with each other and with a microtoroid resonator. The top and middle panels are obtained when side wall is not used to support the fiber tapers. When the microtoroid is brought into focus, only one of the fiber taper seems to be in the same plane with the resonator. When the other fiber is brought to the focus, both the other fiber taper and the resonator are out of focus, revealing the height difference of the double tapers. When the side wall is used, both tapers are brought to the same plane with the resonator and the height difference between them is minimized, if not eliminated, as seen in the bottom panel.

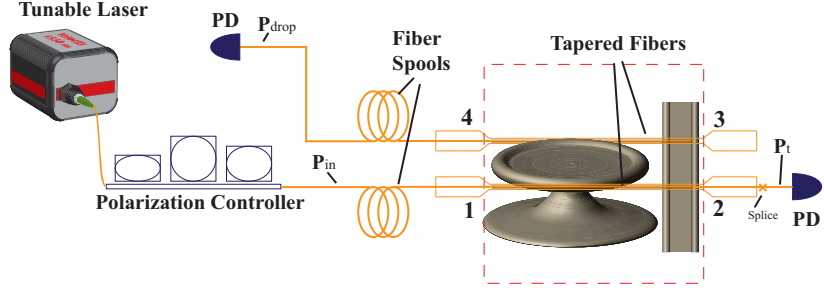

Fig. 7. Experimental setup used for the characterization of the add-drop filter composed a microtoroid resonator coupled to two fiber tapers. The same photodiode (PD) was used to take measurements at the indicated positions. $P_{drop}$: power at drop port; $P_{in}$: input power; $P_t$: transmitted power to the throughput port.

### D. Experimental setup

The setup used in the experiments is depicted in Fig.7. An external-cavity, tunable laser in the wavelength band around 1550nm was used to study and characterize the fabricated ADF. By applying a triangular waveform of period $T$ with amplitude in the range of $0-5V$, the laser frequency (wavelength) can be scanned up and down linearly by 30GHz (0.24nm). In our experiments, the period of the triangular signal was chosen as $T = 16.66$ms and the voltage was in the range of $0-4.2V$, allowing a scanning range of 25.2GHz (0.20nm). The polarization of the input light into the resonator was adjusted by fiber polarization controller to maximize the coupling into the resonator.

The microtoroid that we used in this experiment was 120$\mu$m in major diameter and 9.6$\mu$m in minor diameter. We had a short wall next to each toroid. This design allows us to adjust the horizontal position of the drop waveguide smoothly and precisely after bringing the tapers and the toroid to the same height. We could adjust the gap between the toroid and each of the fiber taper waveguides from under-coupling regime to over-coupling regime. Experiments were performed in the over-coupling regime, thus the gap between the microtoroid and each of the waveguides were within one micron.

As seen in Fig.7, only the add (port 3) and throughput (port 2) ports of the fiber taper waveguides rest on the sidewall. Thus, each of these ports experience ~ 1% loss, while the drop (port 4) and the input (port 1) ports do not. The observed drop efficiency is calculated by taking the ratio of the powers measured at the drop port $P_{drop}$ and the input port $P_{in}$ as

$$D = \frac{P_{drop}}{P_{in}}. \tag{24}$$

We define the cross-talk between the drop and the throughput as the ratio of the transmitted power at the throughput to the dropped power at the resonance frequency, and it is given by $P_t/P_{drop}$.

### E. Channel dropping and loading curve

After aligning the height of the fibers and the microtoroid resonator, we first measured the intrinsic quality factor of the resonator of interest. In order to do this, we moved the add-drop waveguide away from the resonator using a fiber tip such that there is no coupling between the resonator and the add/drop waveguide. Then we set the position of the bus waveguide into the under-coupling regime where $\kappa_0 \gg \kappa_1$. This ensured that the measured quality factor best reflects the intrinsic quality factor $Q_0$. We estimated $Q_0$ of the toroid as $Q_0 = 2.5 \times 10^7$. Next, we tested the dropping behavior of the device. For this purpose, we used the fiber tip again to optimize the distance between the resonator and the fiber tapers such that maximum drop efficiency, minimum cross-talk (i.e., minimum transmission to throughput port) and maximum ADF quality factor $Q$ are obtained. As was discussed in the previous section, not all of these can be satisfied at the same time because the conditions to achieve the first two contradict with the third.

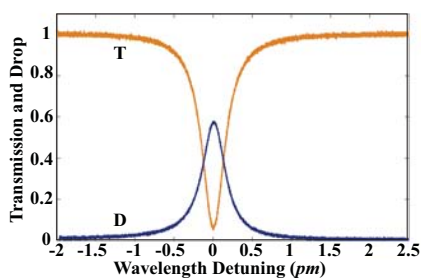

Fig. 8. Oscilloscope screen image captured for the fabricated add-drop filter. The resonance peak and dip denotes the signals for the drop (D) nd throughput (T) ports, respectively. The horizontal signal denotes the transmitted power to the throughput port when the signal is off resonant.

Discussions in the previous section imply that maximum drop efficiency is achieved if both waveguides are in the deep over-coupling regime. Thus, after measuring $Q_0$, we brought the bus-resonator coupling into the over-coupling regime. Then the add-drop waveguide is brought closer to the resonator. At this point we fine-tuned the coupling between the waveguides and the resonator such that maximum signal was obtained at the drop port while the quality factor was still high enough. Figure 8 shows typical signals obtained at the drop and throughput ports as the wavelength of the laser is scanned. At the resonance, a dip and a peak are clearly seen for the throughput and drop ports, respectively. Using the resonance peak or dip in Fig. 8, we estimate the $Q$ of the fabricated ADF as $Q = 4.5 \times 10^6$. Then we moved the add-drop waveguide away from the resonator until it had no effect on the system, leaving only the resonator-bus coupling. The loaded quality factor was measured as $Q_L = 7.9 \times 10^6$ from which we calculate the resonator-bus coupling quality factor as $Q_1 = 1.17 \times 10^7$ using the relation $1/Q_L = 1/Q_0 + 1/Q_1$. Subsequently, we calculate $Q_2$ using the relation in Eq.14 as $Q_2 = 1.04 \times 10^7$. Inserting these values into Eq.10, we estimate the drop efficiency as $D \sim 0.66$. Note that $D$ estimated in this way does not take the transmission losses in the fiber spools, connectors' losses and the detector efficiencies into account. Moreover, the estimated quality factors (i.e., coupling losses and the resonator intrinsic loss) satisfy neither $\kappa_1 = \kappa_0 + \kappa_2$ nor $\kappa_2 = \kappa_0 + \kappa_1$ derived in the previous section for maximal drop efficiency due to the taper's imperfection. We measured the taper transmission to be around 90% and estimated the drop efficiency in our system as $D \sim 0.59$ taking into account the taper loss. A straightforward calculation reveals that if we further decrease $Q_1$ to 63% of its current value by bringing the bus waveguide to the deep over-coupling region $\kappa_1 = \kappa_0 + \kappa_2$ could be satisfied; however, this would sacrifice the $Q$ of the ADF device decreasing $Q$ to 81% of its current value. Similarly, we could increase $Q_2$ by increasing the gap between the add-drop waveguide and the resonator such that the system moves from the deep over-coupling region to a less over-coupled region. In this way, $Q_2$ could be adjusted to be around $2.24 \times 10^7$ so $\kappa_1 = \kappa_0 + \kappa_2$ is satisfied. However, this would not sacrifice $Q$, on the contrary, it would increase $Q$ to $Q \sim 5.9 \times 10^6$.

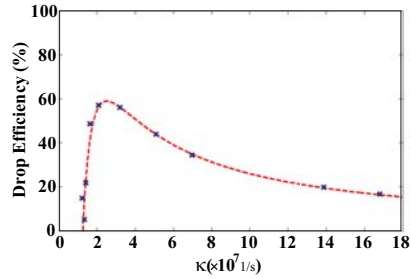

Fig. 9. Experimentally obtained drop efficiency as a function of the total loss $\kappa = \kappa_1 + \kappa_0 + \kappa_2$ of the system and the best fitting curve. $\kappa_0$, $\kappa_1$ and $\kappa_2$ denote the intrinsic resonator loss, bus waveguide-resonator coupling loss and add-drop waveguide-resonator coupling loss, respectively.

Fig.8 shows, transmission (signal at the throughput port) does not reach to zero as the condition $\kappa_1 = \kappa_0 + \kappa_2$ is not achieved. From the power of the throughput port when the laser was on resonance with the cavity mode (i.e., $\Delta = 0$) and when it was off-resonant (i.e., $\Delta \neq 0$, and throughput power equals to input power $P$in), we find $T = 0.0503$. For the drop port, we find $P_{drop} = 0.57P(in)$ at resonance ($\Delta = 0$), corresponding to a drop efficiency of $D = 0.57$. This is slightly less than the drop efficiency of $D \sim 0.59$ which is estimated using the experimentally obtained quality factors in Eq.10 and considering 10% loss in each taper. We attribute this slight difference to the fiber and splicing losses which are not taken into account in Eq.10. Similarly, we find the crosstalk as $T=D = 0.1$ which is at an acceptable level when compared with the reported values for ADFs. It is worth to mention here once more that reducing intrinsic loss $k_0$ (i.e., increasing intrinsic quality factor $Q_0$) of the cavity dramatically increases the drop efficiency. For example, neglecting the taper, connector and fiber losses and keeping $\kappa_1$ and $\kappa_2$ constant as observed in our experiments, just by increasing $Q_0$ to $10^8$ from its current value of $2.5 \times 10^7$ will increase the drop efficiency to 90%.

In order to obtain a loading curve for the fabricated ADF, we calculated drop efficiency *D* while the total loss defined as $\kappa = \kappa_1 + \kappa_0 + \kappa_2$ (or similarly $1/Q = 1/Q_0 + 1/Q_1 + 1/Q_2$) is continuously adjusted by changing the distance between the add-drop waveguide and the resonator (i.e., $\kappa_2$ or similarly $Q_2$ is changed). Decreasing the gap between the add-drop waveguide and the resonator, increases $\kappa_2$ and hence $\kappa$ increases since $\kappa_0$ and $\kappa_1$ are kept intact. The result of these experiments is depicted in Fig. 9 in which we see that increasing $\kappa$ at first increases *D* until it reaches a maximum; however, further increase after this leads a reduction in *D*. The region before the maximum *D* is achieved (region to the left of maximum *D* in Fig. 9) corresponds to translating the coupling between the add-drop waveguide and the resonator from the deepundercoupling region to close to critical coupling, whereas the second region beyond the maximum *D* (region to the right of maximum *D* in Fig. 9) corresponds to transition from the critical-coupling regime to deep-overcoupling. This experimentally obtained drop efficiency dependence on the losses in the system is similar to the theoretical one given in Fig. 3. The observation in these experiments suggests that one can maximize the drop efficiency by properly adjusting the coupling losses in the system.

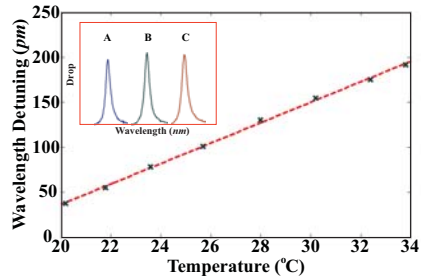

Fig. 10. Thermal tuning of the resonance frequency of a microtoroid and hence the drop frequency of the add-drop filter fabricated using this microtoroid. The inset shows the drop signal at (A) $21.8^{o}C$, (B) $25.7^{o}C$, and (C) $30.2^{o}C$.

### F. Thermal tuning of the drop frequency and thermal stability of drop efficiency

It is known that the resonance frequency of a high-*Q* resonator changes significantly due to thermo-optic effects when the temperature of the resonator or the surrounding medium changes. This property has long been used to fabricate high resolution temperature sensors and to thermally tune resonator-based devices [23]–[25]. On the other hand, the strong thermal response of a resonator may not be acceptable in some applications and solutions to compensate thermal effects are needed [26]. Previously, Little *et al.* [27] and Nawrocka *et al.* [28] have studied the effects of thermal tuning on ADFs fabricated from Hydex and silicon microring resonators, respectively, demonstrating that it is possible to thermally tune the drop wavelength without significantly affecting the filter shape. A similar study has not been carried out for ADFs fabricated from fiber-taper coupled microtoroid or microsphere resonators.

We have performed experiments to investigate the thermal response of the fabricated silica microtoroid based ADFs with and without sidewall. First, we studied the possibility of thermally tuning the resonance wavelength of the device. Thus, we placed the chip on a thermal-electro cooler (TEC) and positioned the bus and drop waveguides as described in previous sections. Then we raised the temperature of the TEC from $20.2^{o}C$ to

33.8°C. In this range, the resonant wavelength linearly increased from 1520nm to 1520.2nm as shown in Fig.10. This suggests that one can tune the drop frequency thermally, but also implies that if the thermal conditions of the ADF are not well-controlled, the drop signal may be different than what is actually intended.

When thermal tuning is to be used to drop different frequencies on-demand, it is crucial that the ADF drops different frequencies without significant deterioration of the drop efficiency. In order to test this for our device, we compared the performance of our device, which has walls for the placements of fiber taper waveguides, with that of a standard device with free-standing fiber-taper waveguides. The results are depicted in Fig. 11. It is seen that for the latter, a change of 1°C reduces $D$ to values less than the half of the initial $D$, and a change of more than 3°C results in zero drop efficiency, implying the deteriorating effect of thermal fluctuation in the designs which use suspended fiber tapers as waveguides in ADFs. Temperature compensation mechanisms should be employed to keep the performance of the ADF intact. Moreover, such an ADF structure would not be suitable for thermally tuned ADF

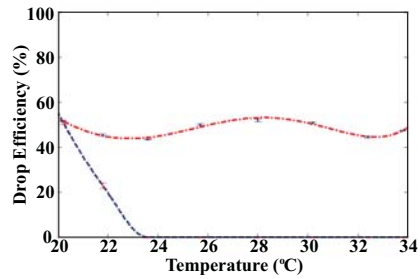

Fig. 11. Comparison of the robustness of drop efficiencies of a microtoroid based add-drop filter with suspended fiber taper waveguides (dashed curve) and the same fiber tapers placed on side walls (dash-dotted curve). For the first, a change of a few degrees is enough to lose the dropping function whereas the drop efficiency for the latter is only slightly affected. The ADF with side wall supporting the fibers demonstrates a much better thermal stability

applications. On the other hand, the new ADF design with side walls for the fiber taper waveguides introduced in this work shows only a little or no change in the drop efficiency as seen in Fig. 11. Within the investigated temperature change of 20.2 − 33.8°C, we observed that $D$ fluctuates in the range of 45 − 55%. This minor change in the drop efficiency of our device with temperature is a substantial improvement over that observed with a chip that did not include a wall. These observations can be explained as follows. When the temperature is increased, we observe from the side view of the resonator that the chip undergoes a significant vertical move, which shifts the resonator upwards while the suspended fiber tapers waveguides stay in their original plane. Consequently, the coupling between the resonator and the fiber tapers gradually deteriorates as the temperature is increased leading eventually to complete loss of the coupling and hence the droping function. If the tapers are situated on the side walls which are fabricated on the same chip as the resonators, a temperature change will shift the chip upwards, which in turn shifts both the resonators and the tapers without significantly affecting their relative positions. Thus, even if the temperature is changed, the coupling between the tapers and the resonators stays the same or changes only slightly. This then provides a robustness to drop efficiency against thermal perturbations. The slight oscillation seen in Fig 11 can be attributed to extra stress induced on the fiber taper due to thermal heating of the side walls and the fiber tapers sitting on them.

IV. CONCLUSION

We have proposed and fabricated an add-drop filter using an on-chip high-*Q* microtoroid with two coupling fiber tapers forming the bus and add-drop waveguides placed on reflowed silica walls placed in close proximity to the microtoroid resonator. We have investigated this new ADF design theoretically and experimentally, clarifying the conditions for maximal drop efficiency and high-*Q* operation of the ADF. We showed that the proposed novel ADF design has a much better thermal robustness and stability than the ADFs with suspended fiber tapers. Moreover, the inclusion of the side walls makes it easier to tune the air gaps between the fiber tapers and the resonator and allows one to precisely placing the waveguides on the same plane with each other and with the resonator. We believe that this work enables an in-depth understanding of the functioning of fiber taper coupled resonators as ADFs and provides a design guideline to construct ADFs with high channel dropping and adding efficiency.